\documentstyle[12pt]{article}

\newcommand{\gO}{\Omega}

\newcommand{\gX}{\Xi}
\newcommand{\vx}{{\bf x}}
\newcommand{\vy}{{\bf y}}
\newcommand{\vz}{{\bf z}}
\newcommand{\vv}{{\bf v}}
\newcommand{\vw}{{\bf w}}
\newcommand{\vnab}{{\bf \nabla}}
\newcommand{\vxi}{{\bf \xi}}
\newcommand{\vgo}{{\bf \omega}}
\title{On the mass transport by a Burgers velocity field}
\author{Paolo Muratore Ginanneschi}
\begin{document}
\maketitle
\centerline{NBI, Blegdamsvej 17 DK 2100, Copenhagen \O, Denmark}
\vfill
\begin{abstract}
The mass transport by a Burgers velocity field is investigated in the 
framework of the theory of stochastic processes. Much attention is devoted to 
the limit of vanishing viscosity (inviscid limit) describing the ``adhesion 
model'' for the early stage of the evolution of the Universe. 
In particular the mathematical foundations for the ansatz currently used in 
the literature to compute the mass distribution in the inviscid limit are 
provided. 
\end{abstract}
\vfill
\centerline{PACS: 05.20, 05.40, 47.20, 47.15}
\vfill
\pagebreak

\section{Introduction}
\label{s:intro}

It was proposed by Zeldovich \cite{Zel}, that a possible model for the 
description of large scale dynamics of the mass distribution, in the early 
stage of evolution of the Universe, is provided by:  
\begin{eqnarray}
\begin{array}{lll}
\partial_{t}\vv +\vv \cdot \vnab \vv =0\,, & \quad
 &  \vv(\vx,t=0) =\vnab V(\vx) \\  
\partial_{t}\rho +\mbox{div}(\vv \rho)=0\,, & \quad 
& \rho(\vx,t=0) =\rho_{0}(\vx) 
\end{array}
\label{Zelmod}
\end{eqnarray}
Here, the mass density is driven by a rotation free velocity field, 
solution of the Riemann equation. 

As it is well known, the Riemann equation with a gradient initial condition 
has multi-stream solutions \cite{GurMolSai},\cite{VeDuFrNo}. 
This phenomenon is easily understood in the Lagrangian picture of the flow: 
the point-mass particles evolving along the characteristics, collide after 
finite time.
The study of the multi-stream solutions requires a very subtle analysis
of the caustics formed after a finite time by the Lagrangian solutions
of the Riemann equation \cite{Arn}.

One way to avoid multistreaming is to introduce the ``adhesion model''
(see \cite{VeDuFrNo}, \cite{Frisch},  \cite{SaiWoy1}, \cite{SaiWoy2} 
with references therein).
\begin{eqnarray}
\begin{array}{lll}
\partial_{t}\vv +\vv \cdot \vnab\vv =\nu\Delta\vv\,, & \quad
 &  \vv(\vx,t=0) = \vnab V(\vx) \\  
\partial_{t}\rho +\mbox{div}(\vv \rho)=0\,, & \quad 
& \rho(\vx,t=0) =\rho_{0}(\vx) 
\end{array}
\label{admod}
\end{eqnarray}
where the velocity field is now governed by the Burgers equation 
\cite{Burgers}.

The introduction of a small diffusion term is expected to have a smoothing 
effect only in the regions where shocks are about to occur: the limit of 
vanishing viscosity of the Burgers equation, the inviscid limit, selects 
one solution for the Riemann equation. 

The main disadvantage of the adhesion model (\ref{admod}) is the loss of
a unique Lagrangian picture for both the velocity field {\em and} the mass 
density. The consequences are not only conceptual but also practical,
since for (\ref{admod}) the solution of the Burgers equation does not 
provide immediately a definite algorithm which solves the continuity equation.
The problem is therefore moved to the mass density evolution.

In the one dimensional case, because of the strong topological constraint,
there is general agreement on the idea that for times long enough that 
shocks appear, an initial uniform density field evolves into a 
singular distribution which describes the formation of point-masses on a 
background of a smooth, diluted, density field. The point-masses are 
situated at the shock positions of the solution to the Burgers equation and
may be referred to as {\em macro-particles}. The mass of a macro-particle is
equal to the integral of the initial mass density extended over the interval
of initial positions which fall into the given shock at some time prior to
that under consideration.

In the two dimensional case the situation seems to be more subtle. Here 
we have the appearance of ruled surfaces, {\em ribbons}, where tangent planes 
touch the graph of the convex hull at a segment instead of a point and of 
{\em triangles}, corresponding to tangent planes with triple contact. In the
Eulerian plane to these regions are respectively associated {\em shock-lines} 
and {\em shock nodes}. 

A further source of complication is that a physically reasonable \cite{SAF} 
initial condition for the Zeldovich model (\ref{Zelmod}) is to assume the 
velocity potential, at time $t=0$, given by a generalized Brownian motion 
in the space variables.

According to the adhesion model (\ref{admod}), in order to compute the mass 
density distribution at any $t$, first, we should integrate the rotation
free Burgers equation and then use the resulting velocity 
$\vv= \vnab \gO$ field to solve the ordinary differential equation 
(ODE)
\begin{equation}
\dot{\vx}_{t}=\vnab \gO(\vx_{t},t\,;\nu) \,,\quad \quad 
\vx_{0}={\bf a}
\label{ode}
\end{equation}
as a function of time and of the initial condition. Then the limit of 
vanishing viscosity should be taken to define the inviscid (weak) limit
of every mass averaged smooth observable.

It is clear that such a program in the realistic three dimensional case is
very resource demanding for its numerical implementation. Therefore the
question naturally arises whether a regularized Lagrangian picture exists 
which simultaneously solves {\em both} the velocity field and the mass 
density equation.
   
In \cite{VeDuFrNo}, Vergassola et al. proposed a numerical algorithm
(the VDFN algorithm from now) based on the generalized Legendre transform 
of the potential field $\gO(\vx,t;\nu=0)$. The underlying ansatz is that
the values of $\vx$, performing the transform, define a solution of 
(\ref{ode}). 

In recent works \cite{SaiWoy1}, \cite{SaiWoy2} it has been proposed that 
a consistent regularization of (\ref{Zelmod}) with a  simple numerical
implementation, can be found by introducing a small viscosity term also in the
right hand side of the continuity equation.

In the present paper it is shown that a Lagrangian picture is uniquely
associated to the Burgers equation. The characteristics of the Burgers equation
are the realizations of the solution of a stochastic differential equation
(SDE) describing a backward diffusion. 

In such a context, the mass transport along the characteristics of the Burgers 
equation is described by a backward Fokker-Planck equation, whose solution
is related to a difficult inverse problem when the given boundary conditions
are the initial velocity potential and the initial mass density distribution. 
(section \ref{sec:Bulag}).

On the other hand the Lagrangian approach shows that the choice of the Burgers 
equation in order to regularize the Riemann equation, is not the natural one 
and it is ``exact'' only in absence of shocks.
In section \ref{sec:forward} the natural regularization is introduced.
It is shown that it provides a simple algorithm in the inviscid limit for
the solution of {\em both} the velocity field and the mass density.
The limitation of such a procedure is that it imposes, in order to be exact, 
some restrictions on the initial conditions of the velocity field. 

In section \ref{sec:Non Lagra} the approaches previously introduced in the
literature (\cite{SaiWoy1}, \cite{SaiWoy2}) are reviewed in light of the
results provided by stochastic calculus. In particular it is shown that
the VDFN algorithm corresponds to an ``effective'' mass transport along
the trajectories of the backward diffusion underlying the Burgers equation.

Furthermore the pair of PDEs specified by the Burgers equation together with 
the ``effective'' mass transport equation, has the Zeldovich model as the 
weak limit for vanishing viscosity for any initial data. The price to pay is 
in the artificial nature of the procedure.

The results of the present analysis suggest that both from the physical and 
the algorithmic point of view, the more natural way to introduce adhesion 
models is that one based on conservation laws \cite{SiRyWe}.

\section{Lagrangian picture for the Burgers equation}
\label{sec:Bulag}

Let us consider the {\em backward} stochastic differential equation
\cite{Guerra}, \cite{GueMo},\cite{GarKon1}, \cite{GarKon2}:
\begin{eqnarray}
d_{s}\vgo_{s}&=&-\vv(\vgo_{s},t-s)\,ds+\sqrt{2\,\nu}\,
d\vw_{s} \quad s\,\le\,t \nonumber \\
\vgo_{0}&=&\vx 
\label{backdrift}
\end{eqnarray}
The Ito differential of the drift field along the flow is:
\begin{eqnarray}
\lefteqn{d_{s}\vv(\vgo_{s},t-s)=}\nonumber\\
&=&[-\partial_{t-s} \vv(\vgo_{s},t-s)-\vv(\vgo_{s},t-s)\cdot \vnab 
\vv(\vgo_{s},t-s)+\nu \Delta \vv(\vgo_{s},t-s)]ds+\nonumber\\
&+&\sqrt{2\,\nu} d\vw_{s}\cdot \vnab \vv(\vgo_{s},t-s)
\end{eqnarray}
where $d\vw_{t}$ defines, as usual, the stochastic differential of Brownian 
motion. 

If $\vv(\vx,t)$ satisfies the Burgers equation, the equality holds:
\begin{equation}
\vv(\vx,t)=<\vv(\vgo_{t}^{\,\vx\,,0},0)>
\label{stochastic}
\end{equation}
where average $<\dots>$ is taken over the realizations of the diffusion
defined by (\ref{backdrift}). 

The physical meaning is clear: the Burgers equation defines a velocity field 
constant on the {\em average} over the random trajectories of 
(\ref{backdrift}). The conservation law allows a straightforward integration 
of the rotation free Burgers equation. 

The basic object to be considered \cite{kara} is the transition probability 
defined by:
\begin{equation}
p_{(-)}(\vy,s\,|\,\vx,t)=<\delta(\vgo_{0}-\vx)\,\delta(\vgo_{t-s}-\vy)> 
\quad s\,\le\,t 
\label{pimeno}
\end{equation}
The transition probability (\ref{pimeno}) satisfies \cite{kara} 
in the variables $(\vy,s)$ the backward Fokker-Planck equation:
\begin{equation}
\partial_{s}p_{(-)} +\mbox{div}(\vv p_{(-)})+\nu\Delta p_{(-)}=0
\end{equation}
with the {\em final} condition:
\begin{equation}
\lim_{s \uparrow t} p_{(-)}(\vy,s\,|\,\vx,t)=\delta(\vy-\vx)
\end{equation}

Let us introduce the auxiliary stochastic process described by the SDE:
\begin{eqnarray}
d\vz_{s}&=&\sqrt{2\,\nu}\,d\vw_{s} \nonumber\\
\vz_{0}&=&\vx
\label{zeta} 
\end{eqnarray}
The transition probability density for this process at time $s$ is easily
found to be Gaussian with expectation value $\vx$ and variance $2\,\nu\,s$.

We can transform (\ref{pimeno}) into an average over the realizations of
$\vz_{t}$:
\begin{equation}
p_{(-)}(\vy,s\,|\,\vx,t)=<\delta(\vz_{0}-\vx)\,
\delta(\vz_{t-s}-\vy)\,\frac{dP_{\vgo_{t-s}}}{dP_{\vz_{t-s}}}
(\vz_{t-s})>
\end{equation}
The Jacobian of this transformation is supplied by the famous Girsanov formula
(see for example \cite{kara} or \cite{GS}): 
\begin{equation}
\frac{dP_{\vgo_{t-s}}}{dP_{\vz_{t-s}}}(\vz_{t-s})=
\exp[-\int_{0}^{t-s}\frac{\ \vnab \gO(\vz_{u},t-u)}{\sqrt{2\,\nu}}
\cdot d\vw_{u}-\int_{0}^{t-s}\frac{\|\ \vnab \gO(\vz_{u},t-u)\|^{2}}
{4\,\nu}\,du]
\label{girfmla}
\end{equation}
where the relation $\vv=\vnab \gO$ has been used. 
Furthermore along the trajectories $\vz_{s}$ the Ito differential of the 
velocity potential becomes: 
\begin{equation}
d_{s}\gO(\vz_{s},t-s)=[-\partial_{t-s}\gO(\vz_{s},t-s)+\nu\,\Delta 
\gO(\vz_{s},t-s)]\,ds+\sqrt{2\,\nu}\,d\vw_{s} \cdot \vnab \gO(\vz_{s},t-s) 
\label{stodiff}
\end{equation} 
If we use (\ref{stodiff}) to eliminate the stochastic integral in 
(\ref{girfmla}) and we impose the normalization condition
\begin{equation}
\int_{-\infty}^{\infty}d^{D}y\,p_{(-)}(\vy,s\,|\,\vx,t)=1
\end{equation}
we find the velocity potential to be
\begin{equation}
\gO(\vx,t)=-2\,\nu\ln\{(\frac{1}{4\,\pi\,\nu\,t})^{\frac{D}{2}}
\int_{-\infty}^{\infty}d^{D}y\,\exp[-\frac{(\vx-\vy)^{2}}{4\,\nu\,t}-
\frac{V(\vy)}{2\,\nu}]\}
\label{velpot}
\end{equation}
and
\begin{equation}
p_{(-)}(\vy,s\,|\,\vx,t)=\frac{\exp[-\frac{(\vy-\vx)^{2}}{4\,\nu\,(t-s)}-
\frac{\gO(\vy,s)}{2\,\nu}]}{\int_{-\infty}^{\infty}\,d^{D}z\,
\exp[-\frac{(\vz-\vx)^{2}}{4\,\nu\,(t-s)}-\frac{\gO(\vz,s)}{2\,\nu}]}
\label{pimenosol}
\end{equation}

It is easy to verify that (\ref{pimenosol}) is a well defined Markovian 
transition probability for any $s\,\le\,t$.

In the physical literature is more common to deal with stochastic calculus
in the Stratonovich representation which has the advantage to preserve
the ordinary rules of differential calculus \cite{Schul}. 

If we express the stochastic Ito integral, appearing in (\ref{girfmla}), 
in terms of the corresponding Stratonovich's \cite{Jona}, then what we have 
done is nothing else than proving that the transition probability 
(\ref{pimeno}) is given by a Feynman path integral with Lagrangian:
\begin{equation}
L(\vx,\dot{\vx},s)=\frac{\|\dot{\vx}+\,\vnab \gO(\vx,t-s)\|^{\,2}}
{4\,\nu}-\frac{1}{2}\Delta \gO(\vx,t-s)
\label{lagrin}
\end{equation}
From (\ref{lagrin}) the derivation of (\ref{pimenosol}) is then trivial.

The solution of the Burgers equation at arbitrary time takes the form:
\begin{equation}
\vv(\vx,t)=\int_{-\infty}^{\infty}d^{D}y\,\vnab V(\vy)\,
p_{(-)}(\vy,0\,|\,\vx,t)
\label{wellknown}
\end{equation}

The knowledge of (\ref{pimeno}) solves also the problem of the
forward evolution of a passive scalar driven by the Burgers equation. 
More explicitly if we consider:
\begin{eqnarray}
\partial_{t}C +\vv \cdot \vnab C &=&\nu\,\Delta C +E \nonumber \\
C(\vx,0)&=&C_{0}(\vx)
\label{concentration}
\end{eqnarray}
where $E=E(\vx,t)$ is an external forcing, then we have 
\begin{equation}
C(\vx,t)=\int_{-\infty}^{\infty}d^{D}y\,C_{0}(\vy)\,p_{(-)}(\vy,0\,|\,
\vx,t)+\int_{0}^{t}\,\int_{-\infty}^{\infty}d^{D}y\,du\,E(\vy,u)
p_{(-)}(\vy,u\,|\,\vx,t)
\end{equation}

As a function of $(\vy,s)$ the transition probability (\ref{pimeno}) 
describes the mass transport along the characteristics (\ref{backdrift}) 
of the Burgers equation.

The equation with final condition
\begin{equation}
\partial_{s}\rho +\mbox{div}(\vv \rho)+\nu\Delta \rho=0\,,  \quad 
\rho(\vx,t=T) =\rho_{T}(\vx) 
\label{rhofin}
\end{equation}
is readily solved for any $t\, \le \,T$ by:
\begin{equation}
\rho(\vx,t)=\int_{-\infty}^{\infty}d^{D}z\,p_{(-)}(\vx,t\,|\,\vz,T)\,\rho_{T}
(\vz)
\label{rhofinsol}
\end{equation}        

On the other hand the boundary conditions for the Zeldovich model provide the
initial mass distribution. In order to solve the mass density evolution in the
interval $[0,T]$, we have, first, to solve the inverse problem:
\begin{equation}
\rho_{0}(\vx)\exp[\frac{V(\vx)}{2\,\nu}]=\int_{-\infty}^{\infty}
\frac{d^{D}z}{(4\,\pi\,\nu\,T)^{\frac{D}{2}}}\exp[-\frac{(\vx-\vz)
^{2}}{4\,\nu\,T}+\frac{\gO(\vz,T)}{2\,\nu}]\,\rho(\vz,T)
\label{massinvers} 
\end{equation}

In principle, once we have solved equation (\ref{massinvers}), it would be
possible to specify the mass density evolution in the interval $[0,T]$. 
The fact that the interval is finite, is not, from this point of view, a 
limitation since $\rho(\vx,T)$ could be consistently used as a new initial 
condition at time $t=T$ to iterate the procedure, as far as the integral 
in (\ref{massinvers}) is convergent.

It is worth to note \cite{GarKon1}, \cite{GarKon2}, that the problem 
(\ref{massinvers}) can be reformulated by looking for the conditions 
that insure the existence of the forward stochastic process specified for 
any $t \in [0,T]$ by the probability density, solution of (\ref{rhofin}) 
with $\rho(\vx,T)$, the unknown final condition to be determined. 

The crucial point is that, in general, the transition probability density 
(\ref{pimeno}) does not have an inverse: because of the non-reversible 
nature of the diffusion dynamics, it satisfies only a semi-group property 
\cite{kara}.

On the other hand, it is interesting to understand the physical meaning for 
the occurrence of the inverse problem (\ref{massinvers}), when our starting 
point is to find a coherent Lagrangian regularization for the Zeldovich model.

\section{The forward diffusion}
\label{sec:forward}

As it is clear from the above discussion, Burgers equation is equivalent to 
the SDE:
\begin{eqnarray}
\begin{array}{lll}
d_{s}\vgo_{s}=-\vv(\vgo_{s},t-s)\,ds+\sqrt{2\,\nu}\,d\vw_{s}\,, 
&\quad \quad & \quad \quad  \vgo_{0}=\vx \\
d_{s}\vv(\vgo_{s},t-s)=\sqrt{2\,\nu}\,d\vw_{s}\cdot \vnab 
\vv(\vgo_{s},t-s)\,, &\quad \quad &  \vv(\vx,0)=\vv_{0}(\vx)
\end{array}
\label{bursde}
\end{eqnarray}
The limit of vanishing viscosity of (\ref{bursde}) is
\begin{eqnarray}
\begin{array}{lll}
d_{s}\vgo_{s}=-\vv(\vgo_{s},t-s)\,ds\,, &\quad \quad &  \quad \quad 
\vgo_{0}=\vx \\
d_{s}\vv(\vgo_{s},t-s)=0\,, &\quad \quad &  \vv(\vx,0)=\vv_{0}(\vx)
\end{array}
\label{burode}
\end{eqnarray}

On the other hand, the characteristic equation of the Zeldovich model 
(\ref{Zelmod}) is
\begin{eqnarray}
\begin{array}{lll}
d_{t}\vxi_{t}=\vv(\vxi_{t},t)\,dt\,, &\quad \quad & \quad \quad \vxi_{0}=\vy\\ 
d_{t}\vv(\vxi_{t},t)=0 &\quad \quad &  \vv({\bf a},0)=\vv_{0}(\vy)
\end{array}
\label{Zeldyn}
\end{eqnarray}
If $\vxi_{t}^{\,\vy,0}$ is a solution of (\ref{Zeldyn}) in $[0,t]$
then $\vgo_{s}^{\,\vx,0}=\vxi_{t-s}^{\,\vx,t}$ is a well defined solution of
(\ref{burode}) {\em only provided} that the solution for (\ref{Zeldyn}) 
exists and is unique in $[0,t]$.

If the latter condition is satisfied, (\ref{burode}), allows to solve the 
inverse problem implicit in the Riemann equation or in the equations for 
any passive scalar conserved along the flow:
\begin{equation}
d_{t}\vv(\vxi_{t}^{\,\vy,0},t)=0 \rightarrow v(\vxi_{t}^{\,\vy,0},t)=
v(\vy,0)
\label{detinvers}
\end{equation} 
Namely, if $\vv$ is well defined for all $s \in [0,t]$ the Riemann equation 
is equivalent to
\begin{equation}
\partial_{s} \vv(\vx,t-s)-\vv(\vx,t-s)\cdot \vnab \vv(\vx,t-s)=0\,, 
\quad \vv(\vx,0)=\vv_{0}(\vx)
\end{equation}
The straightforward consequence is the well known fact \cite{GurMolSai}
that for any $t$ such that no collision occurs between the Lagrangian 
particles, we have:
\begin{eqnarray}
\vv(\vx,t)&=&\vv_{0}(\vgo_{t}^{\,\vxi_{t}^{\,\vy,0},0}) \nonumber\\
\vgo_{t}^{\,\vxi_{t}^{\,\vy,0},0}&=&\vx-\vv_{0}(\vy)\,t 
\label{deterministic}
\end{eqnarray}

It is worth to note that (\ref{deterministic}) and the steps leading to it
are exactly the deterministic counterpart of (\ref{stochastic}) and of the 
procedure providing the integration of the Burgers equation in the particular
case of rotation free initial conditions.

Beside the velocity field, the Zeldovich model requires the mass density field.
It is clear already from the deterministic case that the knowledge of 
$\vgo_{t}^{\,\vx,0}$, alone, does not provide the solution of the continuity 
equation:
\begin{equation}
\partial_{t}\rho +\mbox{div}(\vv \rho)=0\,, \quad  
\rho(\vx,t=0)=\rho_{0}(\vx) \nonumber 
\label{massd}
\end{equation}
The only expression consistent with (\ref{massd}) we can construct, is
\begin{equation}
\rho(\vy,t-s)=\int_{-\infty}^{\infty}d^{D}x\,\delta(\vy-\vgo_{t-s}^{\,\vx,0}) 
\rho(\vx,t)
\end{equation}
where $\rho(\vx,t)$ is a given solution of the continuity equation for 
$t \geq s$.
Again the deterministic case is a mirror of the situation we faced studying
the mass transport by the Burgers equation, i.e. equation (\ref{massinvers}).

The choice to regularize the Riemann equation by resorting to the Burgers 
equation is equivalent to select a solution for the Riemann's, when shocks 
occur, by extending the time reversal conjugation between the exact equation
for the characteristics (\ref{Zeldyn}) and the deterministic limit of 
(\ref{bursde}).
This procedure is of clear advantage when dealing with the velocity field 
alone, but the price to pay consists in the difficulties which arise when 
turning to the problem of the mass evolution.

The alternative approach is to try to regularize the Riemann equation
by looking to the stochastic generalization of its Lagrangian picture.

The SDE to be considered is then:
\begin{eqnarray}
\begin{array}{lll}
d_{t}\vxi_{t}=\vv(\vxi_{t},t)\,dt+\sqrt{2\,\nu}\,d\vw_{t}\,, 
&\quad \quad &  \quad \quad \vxi_{0}=\vy \\
d_{t}\vv(\vxi_{t},t)=\sqrt{2\,\nu}\, d\vw_{t}\cdot \vnab \vv(\vxi_{t},t)\,, 
&\quad \quad & \vv(\vx,0)=\vv_{0}(\vx)
\end{array}
\label{riesde}
\end{eqnarray}
Hence, the system of PDEs implied by (\ref{riesde}) is
\begin{eqnarray}
\begin{array}{lll} 
\partial_{t} \vv +\vv \cdot \vnab \vv +\nu \Delta \vv=0 \,,  &\quad \quad &  
\vv(\vx,t=0)=\vnab V(\vx) \nonumber\\
\partial_{t}\rho +\mbox{div}(\vv \rho)=\nu \Delta \rho \,, &\quad \quad &
\rho(\vx,t=0)=\rho_{0}(\vx) 
\end{array}
\label{avanti}
\end{eqnarray}

For a rotation free initial velocity field, the integration of the pair 
of equations (\ref{avanti}) is achieved in terms of the transition probability
density defined by (\ref{riesde}), following the same steps leading to
(\ref{pimenosol}).

The result is, for any $s\,\leq\,t$:
\begin{equation}
p_{(+)}(\vx,t\,|\,\vy,s)=\frac{\exp[-\frac{(\vx-\vy)^{2}}{4\,\nu\,(t-s)}+
\frac{\gX(\vx,t)}{2\,\nu}]}{\int_{-\infty}^{\infty}d^{D}z\,
\exp[-\frac{(\vz-\vy)^{2}}{4\,\nu\,(t-s)}+\frac{\gX(\vz,t)}{2\,\nu}]}
\label{avantisol}
\end{equation}
where the velocity potential is fixed by the normalization condition and is 
equal to 
\begin{equation}
\gX(\vx,t)=2\,\nu\log\{[\frac{1}{4\,\pi\,\nu\,(T-t)}]^{\frac{D}{2}}
\int_{-\infty}^{\infty}d^{D}z\,\exp[-\frac{(\vx-\vz)^{2}}{4\,\nu\,(T-t)}+
\frac{\gX(\vz,T)}{2\,\nu}]\}
\label{scalavanti}
\end{equation}
Therefore, in order to achieve the solution in the interval $[0,T]$ we need 
to solve the inverse problem:
\begin{equation}
\exp[\frac{V(\vx)}{2\,\nu}]=\int_{-\infty}^{\infty}\,\frac{d^{D}z}
{(4\,\pi\,\nu\,T)^{\frac{D}{2}}}\exp[-\frac{(\vx-\vz)^{2}}{4\,\nu\,T}+
\frac{\gX(\vz,T)}{2\,\nu}]
\label{inversavanti}
\end{equation}
Equation (\ref{inversavanti}) is the natural stochastic generalization of 
the inverse problem (\ref{detinvers}). Once $\gX(\vx,T)$ is known, it can also
be assumed as the new initial condition at time $T$ to iterate the procedure.

In the limit of vanishing viscosity, equation (\ref{inversavanti}) takes 
the form:
\begin{equation}
V(\vx)=\sup_{\vz}[\,\gX(\vz,T)-\frac{(\vz -\vx)^{2}}{2\,T}]
\label{leg1}
\end{equation}
The latter expression is basically a (generalized) Legendre transform 
\cite{VeDuFrNo}:
\begin{equation}
T\,V(\vx)-\frac{(\vx)^{2}}{2}=\sup_{\vz}[T\,\gX(\vz,T)-
\frac{\vz^{2}}{2}+\vx \cdot \vz] \nonumber
\end{equation}
It is tempting to solve (\ref{leg1}) as:
\begin{equation}
\gX\,'(\vx,T)=\inf_{\vz}[V(\vz)+\frac{(\vz -\vx)^{2}}{2\,T}]
\label{solleg1}
\end{equation}
This solution is not exact for a general initial condition: due to the 
convexity properties of the Legendre transform, when we substitute 
$\gX(\vx,T)$ back in (\ref{leg1}) what we find is not $V(\vx)$ 
but its convex hull. Only for this latter the transform (\ref{leg1}) is 
involutive. 
Therefore, by assuming $\gX(\vx,T)=\gX\,'(\vx,T)$ we implicitly adopt the
consistent initial condition.

If we accept, despite of the above restrictions, the pair (\ref{avanti}) as a 
regularization for the Zeldovich model then we are provided a simple algorithm
for the computation in the inviscid limit of both the velocity field, by means 
of (\ref{scalavanti}), and the mass distribution, by means of:
\begin{equation}
\rho(\vx,t)=\int_{-\infty}^{\infty}d^{D}y\,p_{(+)}(\vx,t\,|\,\vy,0)\,
\rho_{0}(\vy)
\label{massavanti}
\end{equation}

\section{Non Lagrangian approaches}
\label{sec:Non Lagra}

In \cite{SaiWoy2} it has been proposed to regularize the Zeldovich model
with the pair of PDEs:
\begin{eqnarray}
\begin{array}{lll}
\partial_{t}\vv +\vv \cdot \vnab\vv =\nu\Delta\vv\,, & \quad
 &  \vv(\vx,t=0) =\vnab V(\vx) \\  
\partial_{t}\rho +\mbox{div}(\vv \rho)=\mu\Delta\rho\,, & \quad
 & \rho(\vx,t=0) =\rho_{0}(\vx) 
\end{array}
\label{massys}
\end{eqnarray}

The meaning of (\ref{massys}) is to assume the velocity field solution
of the Burgers equation as an external field in the Langevin equation 
describing the motion of a point mass particle:
\begin{equation}
d_{t}\vx_{t}=\vv(\vx_{t},t)\,dt+\sqrt{2\,\mu}\,d\vw_{t}\,, 
\quad  \vx_{0}=\vy
\label{massde}
\end{equation}
Hence, the acceleration felt by the point-mass particle is
\begin{equation}
d_{t}\vv(\vx_{t},t)=(\mu+\nu)\Delta \vv(\vx_{t},t)\,dt+
\sqrt{2\,\mu}\, d\vw_{t}\cdot \vnab \vv(\vx_{t},t)
\label{acc}
\end{equation}

It is worth to note that (\ref{acc}) predicts for the case $\mu=\nu$ a 
a mass dynamics different from (\ref{riesde}) even in the limit of vanishing
viscosity: only in absence of shocks the two dynamics become equivalent.

In the general case the solution of mass distribution resulting from 
(\ref{massys}) is difficult. 
The path integral approach clearly shows that it is equivalent to an 
Eu\-clid\-e\-an Schr\"o\-ding\-er equation in a potential given by the 
Laplacian of the velocity potential of the Burgers equation. Namely,
the transition probability for the mass density in (\ref{massys}) is
\begin{equation}
p(\vx,t\,|\,\vy,0)=e^{\frac{\gO(\vx,t)}{2\,\mu}}\,K(\vx,t\,;\,\vy,0)
\,e^{-\frac{\gO(\vy,0)}{2\,\mu}}
\label{mastrans}
\end{equation}
where $K$ is given by
\begin{equation}
K(\vx,t\,;\,\vy,0)=\int_{\vx_{0}=\vy}^{\vx_{t}=\vx} {\cal D}x_{u}\,
\exp[-\int_{0}^{t}(\frac{\dot{\vx}_{u}^{\,2}}{4\,\mu}+\frac{\mu+\nu}{2\,\mu}
\Delta \gO(\vx_{u},u))\,du]
\label{path}
\end{equation}
and it satisfies
\begin{equation}
\partial_{t}K +\frac{\mu+\nu}{2\,\mu}\, K\,\Delta \gO=\mu\Delta K 
\end{equation} 
\begin{equation}
\lim_{t \to 0} K(\vx,t\,;\,\vy,0)= \delta(\vx-\vy)
\end{equation}

There are two evident cases when the path integral (\ref{path}) is of 
practical use. When the Laplacian of the velocity potential does not depend 
on the trajectory, all the integrations turn out to be Gaussian. 
This condition is satisfied for: 
\begin{enumerate}
\item $V(\vx)= \vv_{0} \cdot \vx$. Then the transition probability is: 
\begin{equation}
p(\vx,t\,|\,\vy,0)=\frac{\exp[-\frac{(\vx-\vy-\vv_{0}\,t)^{2}}{4\,\mu\,t}]}
{(4\,\mu\,\pi\,t)^{D/2}}
\end{equation}
which, for every smooth enough initial mass distribution, gives for $\mu
\downarrow 0$: 
\begin{equation}
\rho(\vx,t)=\rho_{0}(\vx-\vv_{0}\,t)
\end{equation}
in accordance to Galilean covariance.
\item $V(\vx)=\frac{\|\vx\|^{2}}{\tau}$. By means of (\ref{path}) we get into
\begin{equation}
p(\vx,t\,|\,\vy,0)=[\frac{\tau}{2\,t+\tau}]^{D/2}\frac{\exp[-\frac{1}
{4\,\mu\,t}\frac{\tau}{\tau+2\,t}(\vx -\frac{\tau+2\,t}{\tau}\,\vy)^{2}]}
{(4\,\mu\,\pi\,t)^{D/2}}
\end{equation} 
In the inviscid limit for smooth enough distribution we obtain:
\begin{equation}
\rho(\vx,t)=(\frac{\tau}{2\,t+\tau})^{D}\rho_{0}(\frac{\tau\,\vx}{2\,t+\tau})
\end{equation}
\end{enumerate}

The peculiarity of these two examples appears also in the fact that the
results do not depend on the viscosity $\nu$ of the driving velocity field. 
Actually the initial conditions for the velocity potential given above, are 
such that no shock appears at any time and the Laplacian of the velocity 
field is zero.

More results can be derived for the case $\mu=\nu$. The crucial observation
is that the fundamental solution of the mass density problem can be  
rewritten as
\begin{eqnarray}
\partial_{t}p +\vv \cdot \vnab p&=&\nu\Delta\, p- p\,\vnab \cdot \vv 
\nonumber\\
\lim_{t \downarrow s} p(\vx,t\,|\,\vy,s)&=&\delta(\vx-\vy)
\label{mastranseq}
\end{eqnarray}
This means that the transition probability $p$ can be computed as an 
average over the random trajectories, solutions of (\ref{bursde})
\begin{equation}
p(\vx,t\,|\,\vy,s)=<J(t,s,\{\vgo \})\,\delta(\vgo_{t-s}^{\,\vx,0}-\vy)> 
\quad s\,\le\,t 
\end{equation}
where
\begin{equation}
J(t,s,\{\vgo \})=\exp[-\int_{0}^{t-s}du\, \vnab \cdot \vv(\vgo_{u}^{\,\vx,0}
,t-u)] 
\end{equation}
is the Jacobian of the change of variables in the functional integration 
between the solutions of the backward diffusion (\ref{bursde}) and those 
of (\ref{massde}).    

In one dimension this observation leads to a straightforward integration
of (\ref{mastranseq}):
\begin{equation}
<J(t,s,\{\omega \})\delta(\omega_{t-s}^{\,x,0}-y)> \equiv 
<\delta(\omega_{t-s}^{\,x,0}-y)\,\partial_{x} \omega_{t-s}^{\,x,0}>
=-\partial_{x} <\theta(y-\omega_{t-s}^{\,x,0})>
\label{gluck}
\end{equation}
where $\theta$ is the step function. It follows:
\begin{equation}
p(x,t\,|\,y,s)=-\partial_{x}\int_{-\infty}^{y}dz\,p_{(-)}(z,s\,|\,x,t)
\label{unod}
\end{equation}
One can easily check that (\ref{unod}) verifies (\ref{mastranseq}) and 
it is a Markovian transition probability.

Unfortunately, the identity in (\ref{gluck}) does not hold in more than 
one dimension.

For the general case, in \cite{SaiWoy2} it has been introduced a ``mean field 
approximation'' for (\ref{mastranseq}). Here a different interpretation
of the result is proposed. 

The mean field theory of \cite{SaiWoy2} is equivalent to substitute to
(\ref{mastranseq}) the equation
\begin{eqnarray}
\partial_{t}p_{*}+\vv \cdot \vnab p_{*}&=&\nu \Delta\,p_{*}-p_{*}\,
<\vnab \cdot\vv> \nonumber\\
\lim_{t \downarrow s} p_{*}(\vx,t\,|\,\vy,s)&=&\delta(\vx-\vy)
\label{effettiva}
\end{eqnarray}
where the average means
\begin{equation}
<\vnab \cdot\vv>=\int_{-\infty}^{\infty}d^{D}x\,\vnab \cdot \vv(\vx,t)
\,p_{*}(\vx,t\,|\,\vy,s)
\label{aver}
\end{equation}
Furthermore we need to impose, for every $t \leq s$, the constraint:
\begin{equation}
\int_{-\infty}^{\infty}d^{D}x\,p_{*}(\vx,t\,|\,\vy,t)=1
\end{equation}

The integration of (\ref{effettiva}) is immediate. Since (\ref{aver})
does not depends on $\vx$, the corresponding term can be extracted from
the path integral, which is reduced to an average over the realizations
of (\ref{bursde}):  
\begin{equation}
p_{*}(\vx,t\,|\,\vy,s)=e^{-\int_{s}^{t}du\,<\vnab \cdot \vv>}
p_{(-)}(\vy,s\,|\,\vx,t) \nonumber
\end{equation}
The normalization condition then fixes the value of the prefactor.
Finally we get into: 
\begin{equation}
p_{*}(\vx,t\,|\,\vy,s)=\frac{\exp[-\frac{(\vx-\vy)^{2}}{4\,\nu\,(t-s)}+
\frac{\gO(\vx,t)}{2\,\nu}]}{\int_{-\infty}^{\infty}d^{D}z\,
\exp[-\frac{(\vz-\vy)^{2}}{4\,\nu\,(t-s)}+\frac{\gO(\vz,t)}{2\,\nu}]}  
\label{Saiapprox}
\end{equation}

Here the velocity potential of the Burgers equation $\gO(\vx,t)$ explicitly
appears. The effect of the average (\ref{aver}) is to define an ``effective'' 
mass transport along the characteristics of the backward diffusion 
(\ref{bursde}). The ``effective'' theory becomes an exact solution
of the continuity equation in the (weak) inviscid limit: for any smooth 
observable described by a scalar function $f(\vx)$, it is readily verified 
that
\begin{equation}
\lim_{\nu \downarrow 0}<f(t\,;\,\nu)>\equiv\lim_{\nu \downarrow 0}
\int_{-\infty}^{\infty}d^{D}x\,d^{D}y\,f(\vx)p_{*}(\vx,t\,|\,\vy,s\,;\,\nu)
\rho_{0}(\vy)
\label{massmed}
\end{equation}
satisfies in the weak sense the continuity equation and we have, 
\begin{eqnarray}
\lim_{\nu \downarrow 0}<f(t\,;\,\nu)>&=&\lim_{\nu \downarrow 0}
\int_{-\infty}^{\infty}d^{D}y\,f[\vx(\vy,t)]\rho_{0}(\vy) \nonumber\\
\vx(\vy,t)&=&\arg \sup_{\vz}[\,\gO(\vz,t\,;\nu=0)-\frac{(\vz -\vy)^{2}}{2\,t}]
\label{ma}
\end{eqnarray}
where $\gO(\vx,t\,;\nu=0)$ is the inviscid limit of velocity potential 
of the Burgers equation.

The latter result (\ref{ma}) is exactly the ansatz of the VDFN algorithm
used in \cite{VeDuFrNo}, in order to compute the evolution of the mass 
density field driven by the Burgers equation. It is also worth to note, that 
in the case of a convex differentiable velocity potential, equation (\ref{ma})
implies free motion for the point-mass particles.

\section{Conclusion}
\label{sec:concl}

The ``natural'' regularization (\ref{avanti}) of the Zeldovich model
by means of the introduction of a small viscosity coefficient leads to the 
inverse problem (\ref{inversavanti}) which is the direct generalization 
of the one occurring in the deterministic case. The inviscid limit drastically 
simplifies the situation, although it imposes some restrictions on the initial 
conditions.
Nevertheless, if we neglect such difficulty, the forward diffusion approach
(\ref{avanti}) provides us a simple algorithm to compute, in the 
inviscid limit, the velocity field {\em and} the mass distribution at any 
time (equations (\ref{scalavanti}),(\ref{solleg1}) and (\ref{massavanti})). 
This procedure is exact for convex initial conditions.

To select the solution of Riemann equation by extending at larger times the  
correspondence with the backward characteristics of the Burgers is of real 
advantage only if we are interested in the velocity field alone.

The occurrence of inverse problems can be avoided if we associate to the 
Burgers equation the ``effective'' equation (\ref{effettiva}) for the mass 
transport along the trajectories of the backward diffusion (\ref{bursde}). 
The solutions (\ref{velpot}) and (\ref{ma}) of the pair of equations 
specified by the Burgers together with its ``effective'' mass transport 
tend in the limit of vanishing viscosity to a weak solution of the Zeldovich 
model.

For the mass density field, such solution (equation (\ref{ma})) was 
proposed as an ansatz in \cite{VeDuFrNo}. The algorithm defined by (\ref{ma}) 
has the advantage of an easy numerical implementation on a computer like 
that one provided by the ``natural'' regularization (\ref{avanti}). 
Furthermore in comparison with the latter it does not impose any restriction 
on the initial conditions.

The main disadvantage of (\ref{ma}) is its intrinsic non-locality which makes 
its use artificial from the microscopic point of view. 

The interpretation of the transition probability (\ref{Saiapprox}), defined by
the ``effective'' theory (\ref{effettiva}), as an approximate solution of the 
exact mass transport (\ref{mastranseq}) by a forward diffusion with external 
drift field given by the Burgers equation, is certainly correct {\em only} 
before the occurrence of shocks. 
For larger times the characteristic equation (\ref{massde}) with $\mu=\nu$ 
seems to indicate a different behavior over shock domains.
Only in one dimension, heuristic arguments can be provided, \cite{SaiWoy2},
to show the equivalence, in the weak inviscid limit, of (\ref{Saiapprox}) with 
the mass transport described by the exact transition probability (\ref{unod}). 

Finally it must be remarked that starting from the basic kinetic equations,
the Zeldovich equations (\ref{Zelmod}) are not the unique starting point
for the construction of adhesion models.
Let us consider 
\begin{equation}
\partial_{t} f + {\bf p} \cdot \ \vnab_{\vx} f =0 
\end{equation}
where $f=f(\vx,{\bf p},t)$. The ansatz
\begin{equation}
f(\vx,{\bf p},t)=\rho(\vx,t)\,\delta[{\bf p}-\vv(\vx,t)]
\end{equation}
leads to the equations
\begin{eqnarray}
\rho \partial_{t}\vv&+&\rho\,(\vv \cdot \vnab) \vv=0
\nonumber\\
\partial_{t} \rho &+& \vnab \cdot (\rho \vv)=0 
\label{Frisch}
\end{eqnarray}
By means of the continuity equation, the system can be recasted in the
form:
\begin{eqnarray}
\partial_{t}(\rho\, v_{i})&+&\vnab_{j} (v_{j}\,\rho\,v_{i}) =0
\nonumber\\
\partial_{t} \rho &+& \vnab_{j} (\rho v_{j})=0 
\label{Sinai}
\end{eqnarray}

In \cite{SiRyWe}, it is proven that the latter pair of PDEs is equivalent to 
the Zeldovich model only in absence of shocks or for an uniform initial
mass distribution.

The simple meaning in terms of conservation laws of equations (\ref{Sinai}) 
allows to achieve an exact algorithm for the solution which turns out
to be of simple numerical implementation \cite{SiRyWe}.

\section*{Acknowledgments}

This work is the result of many discussions with Erik Aurell whom I have
to thank for precious explanations and warm hospitality.
I also benefited from many explanations and comments by Yuri G. Rykov, 
Andrei Sobolevsky, Bualem Djehiche and  Gabriele Travaglini Battino 
Vittorelli.

This work was supported by a 1996 TAO-grant from the European Science 
Foundation. 

The revision of the paper has been done during my stay at the 
CATS group at Niels Bohr Institute with the support of the TMR grant   
ERB4001GT962476 by the European Commission. 

I thank Mogens H. Jensen and all the other cats for useful discussions. 
During the revision of the paper I also benefited from some comments by Piotr 
Garbaczewski which I am pleased to acknowledge.

\end{document}